\documentclass{article}

\usepackage{arxiv}

\usepackage[utf8]{inputenc} 
\usepackage[T1]{fontenc}    
\usepackage{hyperref}       
\usepackage{url}            
\usepackage{booktabs}       
\usepackage{amsfonts}       
\usepackage{nicefrac}       
\usepackage{microtype}      
\usepackage{lipsum}
\usepackage{xcolor}
\usepackage{float}
\usepackage{amssymb}
\usepackage{graphicx}

\title{Empirically Classifying Network Mechanisms}

\author{
    Ryan E.~Langendorf \thanks{We thank Aaron Clauset, Debra Goldberg, Tara Ippolito, Rudy Kahsar, Renae Marshall, and Dan Doak for helpful feedback. This work was funded by the University of Colorado Boulder. Formatting is based on https://github.com/kourgeorge/arxiv-style.}\\
    \footnotesize Cooperative Institute for Research in Environmental Sciences\\
    University of Colorado Boulder\\
    \footnotesize Boulder, CO 80309\\
    \footnotesize \texttt{ryan.langendorf@colorado.edu}\\
\And
    Matthew G.~Burgess\\
    \footnotesize Cooperative Institute for Research in Environmental Sciences\\
    \footnotesize Environmental Studies Program\\
    \footnotesize Department of Economics\\
    \footnotesize University of Colorado Boulder\\
    \footnotesize Boulder, CO 80309\\
}

\begin{document}
\maketitle

\begin{abstract}
Network models are used to study interconnected systems across many physical, biological, and social disciplines. Such models often assume a particular network-generating mechanism, which when fit to data produces estimates of mechanism-specific parameters that describe how systems function. For instance, a social network model might assume new individuals connect to others with probability proportional to their number of pre-existing connections (`preferential attachment'), and then estimate the disparity in interactions between famous and obscure individuals with similar qualifications. However, without a means of testing the relevance of the assumed mechanism, conclusions from such models could be misleading. Here we introduce a simple empirical approach which can mechanistically classify arbitrary network data. Our approach compares empirical networks to model networks from a user-provided candidate set of mechanisms, and classifies each network—with high accuracy—as originating from either one of the mechanisms or none of them. We tested 373 empirical networks against five of the most widely studied network mechanisms and found that most (228) were unlike any of these mechanisms. This raises the possibility that some empirical networks arise from mixtures of mechanisms. We show that mixtures are often unidentifiable because different mixtures can produce functionally equivalent networks. In such systems, which are governed by multiple mechanisms, our approach can still accurately predict out-of-sample functional properties.
\end{abstract}


\section*{Introduction}

Interventions in complex systems and forecasts of their behavior are most likely to succeed under novel conditions when they are based on mechanistic explanations\cite{meinshausen2016methods}. Network data and models describing complex systems have become legion, but there are still relatively few methods for discovering governing mechanisms from network data with statistical tests \cite{broido2019scale} or machine learning \cite{middendorf2005inferring}. Empirically understanding how networks function is increasingly important as scientists are being asked to develop systemic interventions ranging from drugs that alter cellular machinery \cite{vinayagam2016controllability} to management plans for ecosystems in a changing climate \cite{mouquet2013extending} to more just social infrastructure \cite{mosleh2017fair}. 

Empirical network studies often assume a particular governing mechanism (model) and then estimate mechanism-specific parameters. For example, Barab\'{a}si and Albert \cite{barabasi1999emergence} famously found that websites link to each other on the internet according to the preferential attachment mechanism with an attachment power of 2.1. The result is an unequally accessible internet where new websites are exponentially less likely to be linked to and discovered. However, alternative mechanisms are rarely considered in such studies despite evidence that multiple mechanisms can produce structurally similar networks \cite{canard2014empirical, broido2019scale}. Presuming a mechanism for network data enables powerful insights but introduces potential for tautological conclusions. Network scientists can avoid this issue by non-parametrically correlating properties of networks \cite{dunne2002network} or individual nodes \cite{vinayagam2016controllability} with important outcomes like stability or persistence, but at the cost of understanding the mechanistic nature of these correlations which is critical for effective intervention.

Here, we introduce a general approach to empirically classify network mechanisms. By comparing an unknown network to networks simulated from known mechanisms we can--with high sensitivity and specificity--classify any empirical network as either resulting from one of a candidate set of mechanisms, or being the product of none of them.

\section*{Network comparisons distinguish mechanisms}

\begin{figure*}[th!]
\centering
\includegraphics[width=\linewidth]{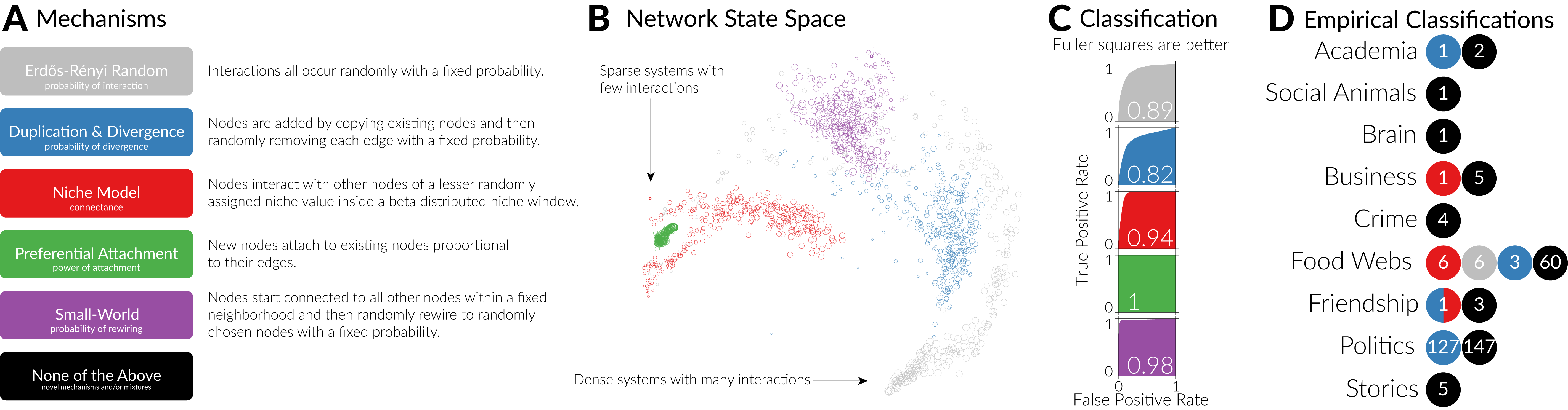}
\caption{\scriptsize The classification of five common network \textbf{Mechanisms}. 100 networks were simulated from each mechanism, systematically varying one governing parameter for each network, with 3 replicates of each value. This parameter is listed below the name of each mechanism respectively. Note that our approach is not limited to these five mechanisms. A \textbf{Network State Space} was made from the 1500 simulated networks by pairwise comparing 18 properties of each network in a stacking ensemble method. A two-dimensional NMDS projection of this state space (with identical axis scaling) shows that networks cluster by mechanism. In this plot circles are each a network, with their radius proportional to the governing parameter. \textbf{Classification} ROC curves and corresponding AUC values (inset numbers) for the ability to classify these 1500 networks. When AUC = 1 the classifier can identify all true positives without including any false positives. Random classifiers produce AUC = 0.5. Each network was classified with an independently simulated state space. \textbf{Empirical Classifications} of 373 empirical networks spanning 9 kinds of systems. Each circle represents a mechanism. Inset numbers are the number of networks classified as that mechanism. If a network was classified as more than one mechanism the resulting pie graph was made with equal splits for each mechanism because we cannot yet confidently assign probabilities to mixture networks as shown in Fig.~\ref{Fig:EmpiricalPies}.}
\label{Fig:Classifier}
\end{figure*}

Our method, which is available in the open-source R package \textit{netcom} \cite{netcom}, is a comparative approach to network classification. It systematically compares a network of interest to networks simulated from candidate mechanisms. Here we construct these candidate networks by growing networks where nodes attach to each other according to the rules of one of the five mechanisms listed and defined in Fig.~\ref{Fig:Classifier}A: Erd\"{o}s-R\'{e}nyi random \cite{erdos1959random}, Duplication and Divergence \cite{ispolatov2005duplication}, the Niche Model \cite{williams2000simple}, scale-free Preferential Attachment \cite{barabasi1999emergence}, and Small-World networks \cite{watts1998collective}. Our approach allows us to test if a network came from any of these five mechanisms, and can readily incorporate any other mechanism that can be simulated.

We use a stacking ensemble approach to capture the many ways network structures differ across mechanisms, combining 9 network properties into a measure of how different two networks are: in and out degrees, entropy of in and out degrees, clustering coefficient (transitivity), Google's PageRank \cite{brin1998pagerank}, the number of communities \cite{clauset2004finding}, and the numbers of each possible 3-node and 4-node motifs. Additionally, we recalculate each of these on a fifth-order row-normalized Markov version of each network to include indirect effects. We combine (stack) these 18 network properties into a single number measuring the difference $d \left( N_i, N_j \right)$ between two networks $N_i$ and $N_j$. To do this, we measure the Euclidean distance between the values of each property in the two networks, and calculate the average across all 18 properties weighted by each properties' loading in the first axis of a principal components analysis of networks simulated systematically across all candidate mechanisms (Supporting Dataset~1).

The set of $d \left( N_i, N_j \right)$'s across all pairs of networks constitutes a network state space (e.g. Fig. \ref{Fig:Classifier}B) within which closer networks have more similar structures, function more similarly, and are more likely to have been generated by the same mechanism. To test if a network came from a particular mechanism our classifier simulates networks from that mechanism, systematically varying the mechanism's parameters to find the version of that mechanism closest to the unknown network. It then simulates networks from this region of the network state space and uses a KS test to determine if the distances between networks from that best fit version of the mechanism likely came from the same distribution as the distances between them and the unknown network. 

We also note that the parameter values within each mechanism (size of each point in Fig.~\ref{Fig:Classifier}B) vary smoothly in the network state space suggesting this approach may be able to both classify and parameterize mechanisms. Our package \textit{netcom} \cite{R, netcom} estimates parameter values using an average of the known network parameters weighted by their distances from the unknown network.

An ROC (Receiver Operating Characteristic) curve judges a classifier by showing the trade-off between true and false positive labels. In the context of inferring network mechanisms, ROC curves quantify how likely a network classified as a particular mechanism is to actually be from that mechanism. The high AUC (Area Under the Curve) values in Fig.~\ref{Fig:Classifier}C indicate that our classifier has both a high true positive rate and a low false positive rate. This confirms our approach can identify when an empirical network was not produced by any of the candidate mechanisms.

\section*{Empirical network classification}

We used our classifier (the function `classify' in \textit{netcom} \cite{netcom}) to test 373 empirical networks spanning a wide range of physical, biological, and social systems (Fig.~\ref{Fig:Classifier}D). All 373 are freely available (for sources, see Supporting Dataset~2).

Almost half (46\%) of the political networks were classified as governed by Duplication and Divergence. This is intuitive: newly elected officials often begin networking similarly to each other. Other kinds of networks were mostly (80\% on average) classified as none of the five considered mechanisms. Among successfully classified networks some were intuitive (e.g. Niche Model food webs), but many were not, and some were inconsistent (e.g. Friendship). Many commonly assumed system-mechanism pairings were not supported by our classifier (e.g. small-world brain networks \cite{bassett2006adaptive}).

Why did so few of these empirical networks classify as one of the five mechanisms? One possibility is that these systems are governed by other mechanisms not considered. A second intriguing possibility is that the unclassified systems are governed by a mixture of multiple mechanisms (e.g. 20\% preferential attachment and 80\% random).

\section*{Mixture mechanism identifiability}
Mixture networks could appear enough like those from a single mechanism to be studied as one, but the mixture would bias parameter estimates. For instance, consider a growing network assumed to be governed by Preferential Attachment, but whose first nodes were actually interacting randomly. Estimates of the preferential power of attachment, which measures inequity in the system, would underestimate how unfair this system actually is. The co-sponsorship network of bills in the US Senate provides a real-world example. Fowler \cite{fowler2006legislative} estimated an attachment power of 6.37--suggesting consensus is rare--but found an attachment power of 1.59 when only considering senators in the same political party. Thus, instead of an overly unfair legislative mechanism governed exclusively by preferential attachment, bill sponsorship may be the result of a more complex and fair intra-party process. Similarly, assuming the internet is governed exclusively by the preferential attachment mechanism, with an estimated attachment power of 2.1 \cite{barabasi1999emergence}, assumes every website links to other websites through an identical process. This is unlikely considering many websites are now made from templates with common media links, acting then in part according to the Duplication and Divergence mechanism. How misleading are these single-mechanism parameter estimates?

\begin{figure*}[th!]
\centering
\includegraphics[width=\linewidth]{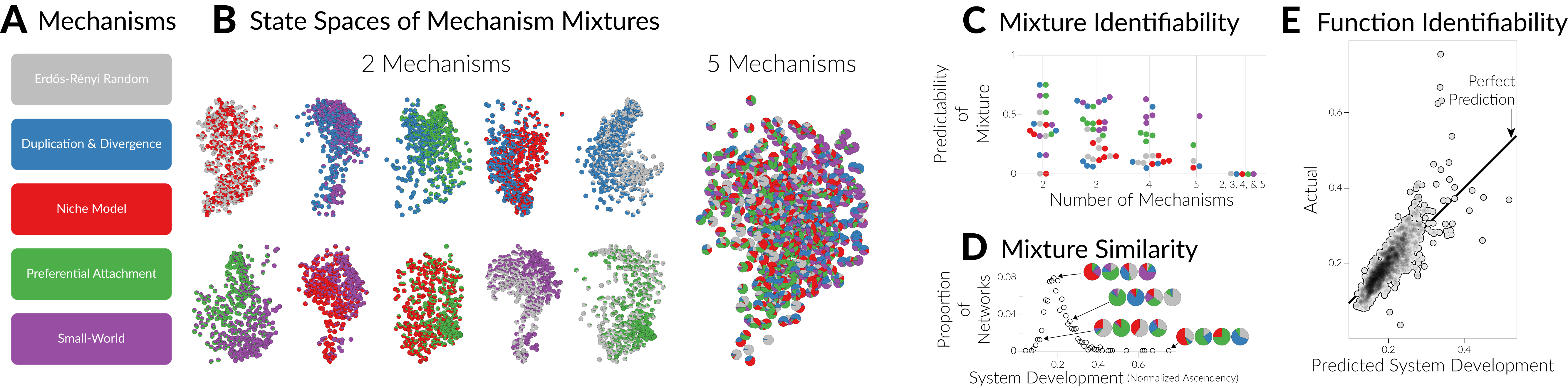}
\caption{\scriptsize Networks governed by mixtures of \textbf{Mechanisms}. NMDS projections (each with identical axis scaling) of the ensemble distances between networks, as in Fig.~\ref{Fig:Classifier}B, show \textbf{State Spaces of Mixture Mechanisms}. Networks (pies) were made with the `grow\_Mixture' algorithm in \cite{netcom}. The colors in each pie are proportional to the number of nodes in that network governed by the corresponding mechanism. \textbf{Mixture Identifiability} Smaller mixtures (e.g. two mechanisms) may be identifiable, but \textit{de novo} identification of mixture mechanisms across larger numbers of mechanisms (e.g. five) or when the number of mechanisms is unknown (far right column) is not possible with our approach. \textbf{Mixture Similarity} Different mixtures of mechanisms can produce functionally equivalent systems, here quantified by normalized Ascendency\cite{ulanowicz2012growth}. Pies are example networks generated by mixtures of the five mechanisms, grouped by their shared Ascendency. \textbf{Function Identifiability} Leave-one-out cross validation of system development (normalized Ascendency). Each point is a single mixture network, with randomly chosen proportions of the five mechanisms and governing parameters within each mechanism.}
\label{Fig:EmpiricalPies}
\end{figure*}

We are unaware of software to simulate mixture networks, which are needed for training, classifying, and ground-truthing mixture network models. To address this our R library \textit{netcom} includes functions that generate mixture networks in two ways: (i) networks are grown one node at a time each of which attaches to existing nodes according to one mechanism (the function `grow\_mixture', used in Fig.~\ref{Fig:EmpiricalPies}); and (ii) starting with a random fully grown network before, in a random order, iteratively rewiring nodes according to fixed node-specific mechanisms (the function `stir\_mixture'). 

We find that mixture mechanisms are less identifiable than pure mechanisms. Fig.~\ref{Fig:EmpiricalPies}C shows that as the number of mechanisms in each network increases from two to five, nearby networks in the state-space no longer have more similar mixtures of mechanisms than far-apart networks. Our approach can classify mixtures of only two known mechanisms (Fig.~\ref{Fig:EmpiricalPies}B,C), but we cannot first justify excluding the other three. 

The reason we cannot classify arbitrary mixtures of network mechanisms is that our approach effectively measures and compares network functioning, and entirely different mixtures can produce networks with identical functioning. Fig.~\ref{Fig:EmpiricalPies}D illustrates this with Ascendency\cite{ulanowicz2012growth}, a thermodynamic measure of system growth and development not included in our classification ensemble. As Fig.~\ref{Fig:EmpiricalPies}E shows, our approach can be used to accurately predict out-of-sample Ascendency of all sizes and proportions of mixture networks. Thus, even on mixture networks, our approach serves its core purpose of classifying networks according to how they function. And, the non-uniqueness of mixture networks is actually an asset to our approach, making it easier to build a set of candidate models that meaningfully span a functional space within which empirical networks can be classified.

\section*{Conclusions}
Knowing the mechanisms that govern a system is key to predicting its behavior in novel conditions. Our results show that we can infer mechanisms from network data by comparing empirical networks to simulated networks across an ensemble of structural properties. We have also shown that more complicated systems governed by different mixtures of mechanisms can produce functionally equivalent networks. While this prevents our approach from inferring the proportions of each mechanism at play in such a network, out approach can still predict the way it functions. These results illustrate the rich mechanistic information carried in network properties, and the utility of comparative inferences, which are becoming more common in network science \cite{sharan2006modeling, bandyopadhyay2006systematic, papin2004comparison, jantzen2016dynamical}. In thinking about interventions, and exogenous disturbances, we cannot rule out the possibility that functionally identical mixtures would functionally diverge following a structurally significant disturbance and subsequent (re-) growth. This possibility, and mixture networks more generally, merit further study.

\bibliographystyle{unsrt}  
\bibliography{ms}

\end{document}